\documentclass[12pt]{iopart}
\usepackage{graphicx}

\newcommand{\be}{\begin{equation}}
\newcommand{\ee}{\end{equation}}
\newcommand{\beq}{\begin{eqnarray}}
\newcommand{\eeq}{\end{eqnarray}}

\begin{document}

\title{Subthreshold $\phi$ meson production in heavy-ion collisions}

\author{M Z\'et\'enyi\dag, H W Barz\ddag, Gy Wolf\dag\ and
B K\"ampfer\ddag}

\address{\dag\ KFKI Research Institute for Particle and Nuclear Physics,
POB 49, H-1525 Budapest, Hungary}

\address{\ddag\ Forschungszentrum Rossendorf, 
Pf 510119, 01314 Dresden, Germany}

\begin{abstract}
Within a BUU type transport model we study $\phi$ meson production 
in subthreshold Ni+Ni and Ru+Ru reactions. 
For the first time we included in our model the elementary reaction 
channels $\rho+N,\Delta \to \phi+N$, $\pi+N(1520) \to \phi+N$
and $\pi \rho$ $\to$ $\phi$. 
In spite of a substantial increase of the $\phi$ multiplicities 
by these channels our results stay significantly below the 
preliminary experimental data. 
\end{abstract}


\section{Introduction}
Meson production is a sensitive probe of the reaction dynamics
of heavy-ion collisions. This is particularly true for the 
production of heavy mesons in subthreshold reactions. 
Open strangeness production 
in the threshold region has been studied extensively both 
experimentally and theoretically. On the contrary,
only a few measurements have been reported on hidden strangeness 
(i.e.\ $\phi$ meson) production. 
Recent preliminary results for the reactions
 Ni + Ni at 1.93 A$\cdot$GeV and
Ru + Ru at 1.69 A$\cdot$GeV by the FOPI collaboration \cite{kotte}
point to a surprisingly large production cross section. 
Earlier transport model calculations \cite{chung2} underestimate
these data substantially.
In those calculations $\phi$ mesons are produced in the elementary
reactions BB $\to$ $\phi$BB, $\pi$B $\to$ $\phi$B and K\=K $\to$ $\phi$, 
where
B stands for a nucleon or a $\Delta$ baryon. 
The discrepancy of the calculations and the data raises the question
of whether further elementary channels might essentially contribute
to the $\phi$ meson yield.

The relatively large branching ratio of the $\phi \to \pi\rho$
decay channel suggests that processes with a 
$\phi\rho\pi$ vertex and hence
the $\rho$ meson can play an important role.
Therefore in this contribution we consider $\phi$ meson production in 
elementary processes with a $\rho$ meson in the entrance channel,
namely $\rho$B $\to$ $\phi$B and $\rho\pi$ $\to$ $\phi$.
We also test the effect of higher resonances by studying the 
elementary reaction $\pi$N(1520) $\to$ N$\phi$ via $\rho$ meson exchange.
The reason for choosing the N(1520) resonance is its large 
partial decay width to $\rho$N.
Diagrams representing these elementary reaction channels
are shown in Fig.\ \ref{diagrams}.

\section{Cross sections of elementary reactions}

\begin{figure}
\begin{center}
\includegraphics[width=6.5cm]{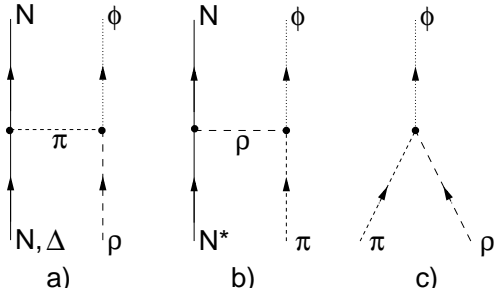}
\end{center}
\caption{Diagrammatic representation of processes contributing to the
reactions (a)
$\rho$ (N,$\Delta$) $\to$ $\phi$ N, (b)
$\pi$ N$^*$ $\to$ $\phi$ N and (c) $\pi \rho$ $\to$ $\phi$.}
\label{diagrams}
\end{figure}

\begin{figure}
\begin{center}
\includegraphics[width=7.2cm]{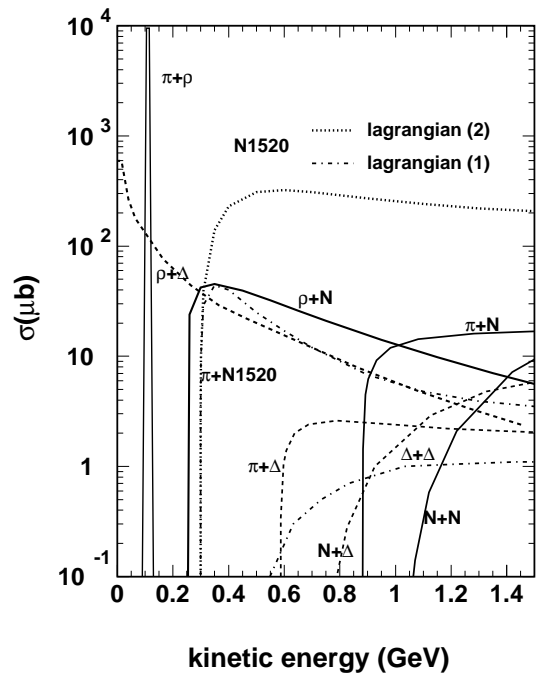}
\end{center}
\caption{Production cross sections for $\phi$ mesons for
several binary reactions as a function of the 
relative kinetic centre-of-mass energy in the entrance channel.
The cross sections are calculated for the nominal masses of
the resonances. Especially cross sections  for $\rho \Delta$ collisions
having different masses can considerably deviate from the curve shown.
}
\label{crosssec1}
\end{figure}

Since the cross sections of the elementary reactions are not well known
experimentally we employ a one-boson exchange model to estimate them. 
For the BB $\to$ BB$\phi$ and the $\pi$B $\to$ $\phi$B channels we use the 
same cross sections as in \cite{chung2}.
The calculations of the $\rho N \to \phi N$ and $\rho\Delta \to \phi N$
cross sections (Fig.\ \ref{diagrams}a) are based on the  effective 
interaction Lagrangians 
\[
{\mathcal L}_{\pi NN} = \frac{f_{\pi NN}}{m_{\pi}}
\bar{\psi}\gamma_5\gamma^{\mu}\vec{\tau}\psi\cdot\partial_{\mu}\vec{\pi},
\;\;\;\;\;\;
{\mathcal L}_{\pi N\Delta} = - \frac{f_{\pi N\Delta}}{m_{\pi}}
\bar{\psi}\vec{T}\psi^{\mu}\cdot\partial_{\mu}\vec{\pi} \;+\; \mbox{h.c.},
\]
\[
{\mathcal L}_{\pi\rho\phi} = \frac{f_{\pi\rho\phi}}{m_{\phi}}
\epsilon_{\mu\nu\alpha\beta}
\partial^{\mu}\phi^{\nu}\partial^{\alpha}\vec{\rho}^{~\beta}\cdot\vec{\pi}.
\]
Here $\psi$ and $\psi_{\mu}$ are the nucleon field and the Rarita-Schwinger 
field for the $\Delta$ resonance, respectively, and $\pi$, $\vec{\rho_{\mu}}$, 
$\phi_{\mu}$ stand for the meson fields.
We include multipole form factors for off-shell pions and $\rho$ mesons.
Coupling constants and cutoff parameters 
are taken from \cite{chung2,machleidt}.
The resulting isospin averaged cross sections are shown
in Fig.\ \ref{crosssec1}.
It should be noted that the cross section of the $\rho$B channels also
depends strongly on the actual mass of the incoming $\rho$ meson
(and $\Delta$ particle).

To calculate the cross section of the $\pi$N(1520) channel 
(Fig.\ \ref{diagrams}b) we tested several possibilities for
the interaction Lagrangian for the $\rho$NN(1520) vertex
such as 
\beq
\label{vec1520}
{\mathcal L}_{\rho NN(1520)} & = & \frac{f_1}{m_\rho}
\bar{\psi}_N \vec{\tau} \gamma ^\nu \psi_{1520}^\mu \,
(\partial_\nu \vec{\rho}_\mu - \partial_\mu \vec{\rho}_\nu) 
\;+\;\mbox{h.c.},\\
\label{ten1520}
{\mathcal L}_{\rho NN(1520)} & = & \frac{f_2}{m_\rho^2}
\bar{\psi}_N \vec{\tau} \sigma^{\alpha\beta} \psi_{1520}^\mu \,
\partial_\alpha \partial_\mu \vec{\rho}_\beta  \;+\;\mbox{h.c.}\;.
\eeq
We fitted the coupling constants to the $\rho$(two-pion) decay channel
of the N(1520) and obtained the values $f_1 = 10.5$ and $f_2 = 68$.
We observe that 
Lagrangian (\ref{ten1520}) gives a much larger cross section for $\phi$ meson
production than Lagrangian (\ref{vec1520}) (see Fig.\ \ref{crosssec1}).
However, the value of the coupling $f_2$ extracted from the N(1520) decay
differs significantly from the value estimated within the quark model.
As the relevant energies in the decay and the production process
are very different the assumption of a constant coupling might fail.
In our BUU calculations reported below
we use the cross sections obtained using Lagrangian
(\ref{vec1520}) as standard. We also carried out some calculations
using Lagrangian (\ref{ten1520}) to test the uncertainties of the $\phi$ 
meson yield from the $\pi$N(1520) channel.

\section{BUU calculations of the $\phi$ meson yield}
We used the above elementary cross sections to calculate the $\phi$
meson production within a BUU model. In our model nucleonic and
$\Delta$ resonances are included up to 2.2 GeV. The resonance properties
are fitted to one- and two-pion production data in $\pi$N scattering 
while the production cross sections of resonances are derived from
meson production data in NN collisions. We employed a soft 
momentum dependent 
potential for nucleons and $\Delta$(1232). The in-medium mass of the
$\phi$ meson is treated according to \cite{hatsuda}. We included also 
the absorption of $\phi$ mesons via the  
$\phi$N$\to$K$\Lambda$ \cite{haglin}
and the $\phi$N elastic scattering with a cross section of 0.56 mb.

\begin{table}[t]
\begin{center}
\begin{tabular}{l|c|c}
 yields from & Ni + Ni (1.93 GeV) & Ru + Ru (1.69 GeV) \\ 
\hline
   B + B           &  $3.5\cdot 10^{-4}$ &  $3.1 \cdot 10^{-4}$ \\
   $\pi$ + B       &  $2.9\cdot 10^{-4}$ &  $3.2 \cdot 10^{-4}$ \\
   $\rho$ + B      &  $8.9\cdot 10^{-4}$ &  $11.8\cdot 10^{-4}$ \\
   $\pi$ + $\rho$  &  $1.6\cdot 10^{-4}$ &  $1.5 \cdot 10^{-4}$ \\
   $\pi$ + N(1520) &  $0.5\cdot 10^{-4}$ &  $0.6 \cdot 10^{-4}$ \\
\hline
 total yield       &  $1.7\cdot 10^{-3}$ &  $2.0 \cdot 10^{-3}$ \\
\hline
experiment \cite{kotte} &  $(8.7\pm3.6)\cdot 10^{-3}$  &  
$(6.4\pm2.5)\cdot 10^{-3}$
\end{tabular}
\end{center}
\caption{Multiplicities of $\phi$ mesons 
per central event in Ni + Ni and Ru + Ru reactions. 
The first 5 lines give the results from the special channels indicated.
The symbol B comprises the nucleon and the $\Delta$ particle.
The cited experimental yields are to be considered very preliminary.}
\label{4piyield}
\end{table}

The resulting $\phi$ meson yields 
are shown in
Table~\ref{4piyield} together with the data obtained by extrapolating
the experimental yields to the entire phase space.
The inclusion of the new $\rho$ channels results in a substantial
increase of the $\phi$ meson yields by a factor of 2.5.
The dominant contribution comes from the $\rho$B channel.
The $\pi$N(1520) channel calculated with the Lagrangian 
(\ref{vec1520}) does not contribute essentially to the $\phi$
meson yield. However, using the Lagrangian (\ref{ten1520}) the yields
of this channel rise by a factor of about 10 
pointing to the importance of this channel for $\phi$ meson production.
The experimental yields are still underestimated by our calculations,
this statement holds independently of the uncertainties of 
the $\pi$N(1520) channel.

\section{Conclusions}
We have presented a calculation of the $\phi$ meson production
in heavy-ion collisions including in our BUU model the 
elementary channels
$\rho$B$\to\phi$N, $\pi$N(1520)$\to\phi$N and $\pi \rho$ $\to$ $\phi$
for the first time. 
As a result the $\phi$
meson yields are substantially increased but the preliminary experimental 
data are still underestimated. A dedicated measurement with enlarged
statistics would be highly welcome to resolve the $\phi$ puzzle.
For a more detailed discussion of our results see \cite{ourpaper}.

\section*{Acknowledgments} 
We thank R.\ Kotte for continuous information on his analyses of the
FOPI data. 
This work was supported in part by the German BMBF 
grant 06DR921, the DAAD scientific exchange program with Hungary,
and the National Fund for Scientific
Research of Hungary, OTKA T30171, T30855 and T32038.


\end{document}